# Lamb wave-based MVDR imaging and CNN classification of defects in pipelines

Shuangshuang Li[1], Kai Zhao[1,2*]

[1] *School of Mechanical Engineering, Jiangnan University, Wuxi 214122, China;*

[2] *Jiangsu Key Laboratory of Advanced Food Manufacturing Equipment Technology, Wuxi 214122, China*

**Abstract**: Significant progress has been made in ultrasonic guided wave (UGW) technology for pipe signal processing and defect imaging recently. However, developing a defect localization and imaging algorithm that requires fewer parameters, offers a wide imaging range, and achieves high positioning accuracy remains a considerable challenge. Traditional direction-of-arrival (DOA) estimation methods primarily focus on the single-angle estimation with low resolution, failing to satisfy the spatial localization requirements for pipeline defects. Therefore, a high-resolution spatial spectrum estimation algorithm is introduced to realize the two-dimensional DOA estimation. By distributing multiple sensors in a specific geometric configuration to form an array, this method employs the array signal processing technology to accurately obtain the DOA of spatial signals. A uniform circular array (UCA) is employed in the present study to receive signals from pipe defects, and high-precision localization and imaging of defects are achieved based on the two-dimensional minimum variance distortionless response (MVDR) beamforming algorithm, with a relative positioning error of less than 1%. An image classification method based on the convolutional neural network (CNN) is further developed to distinguish the defect types. By constructing a novel CNN model to extract defect features and perform classification, this model achieves a prediction accuracy of 97.50%, which effectively distinguishes between defect types.

## 1. Introduction

As the essential component of modern infrastructures, pipelines play an irreplaceable role in the transportation of oil and gas. Due to the long distances covered by these pipelines, they are prone to damages such as cracks, holes, and corrosion during operation under the influence of complex service environments, leading to the leakage of fluids inside pipelines. If not detected and repaired promptly, these defects may cause catastrophic failures, posing severe threats to public safety and resulting in substantial economic losses [1]. Therefore, the pipeline damage monitoring has become an essential requirement for ensuring energy transportation security. Ultrasonic Lamb waves are widely used in the fields of non-destructive testing (NDT) and structural health monitoring (SHM) of complex structures such as pipelines [2], composite material plates [3], and aircraft blades [4] because of their high penetration capability, high sensitivity to defects, and long-distance propagation characteristics. However, due to the multi-modal and dispersive characteristics of Lamb waves, traditional imaging and positioning methods face technical bottlenecks such as difficult signal decoupling and insufficient positioning accuracy under sparse sensor configurations [5-7].

* Corresponding author. *E-mail addresses*: kai.zhao@jiangnan.edu.cn (K. Zhao).

With the widespread usage of millimeter-wave multiple-input multiple-output (MIMO) systems in wireless communication and microphone arrays in sound source localization, the direction-of-arrival (DOA) estimation of spatial signals has been attracting considerable attentions as an advanced array signal processing technology [8, 9]. Specifically, the DOA method operates by receiving spatial signals through sensor arrays and performing super-resolution estimation of the signal arrival directions. Within the DOA estimation framework, the azimuth estimation based on one-dimensional parameters is defined as one-dimensional DOA estimation, while the two-parameters estimation combining azimuth and elevation angles is called two-dimensional DOA estimation [10]. In terms of the structure of sensor arrays, compared with configurations such as the L-shaped array, uniform linear array, and cross-shaped array, the uniform circular array (UCA) provides omnidirectional coverage, has higher directional resolution in the circular plane, and supports the joint estimation of azimuth and elevation angles [11]. In addition, since the array elements are uniformly distributed along the circumference, the UCA has higher sensitivity to changes of the signal source angle, thus possesses considerable advantages in pipe defect detection. The excellent target directivity and anti-interference ability of the DOA estimation already make it widely used in numerous economic fields, such as speech enhancement [12], noise suppression [13], wireless communication [14], unmanned aerial vehicle swarms [15], underwater acoustics [16], and seismology [17].

DOA estimation methods generally encompass beamforming algorithms, subspace-based algorithms, and sparse signal processing algorithms [18, 19]. Beamforming algorithms mainly include the conventional beamforming (CBF) [20] and minimum variance distortionless response (MVDR) beamforming [21] algorithms. Although the CBF has advantages such as simple implementation, low computational complexity, and stable performance, it also usually struggles to meet the application requirements of high-precision beamforming due to high imaging side lobes and weak noise suppression ability. Fortunately, the MVDR algorithm could effectively overcome the above shortcomings. By optimizing the weighting vector, it suppresses the side lobe level while maintaining the main lobe width, breaking through the Rayleigh resolution limit of the conventional beamforming. This algorithm could effectively enhance useful signals while suppressing useless interference and noise signals [22-24]. Nevertheless, this approach still exhibits certain constraints, including reduced target resolution under low signal-to-noise ratio (SNR) conditions and high computational demands [25]. Zhu *et al*. [26] proposed an enhanced beam control algorithm based on MVDR for a multi-channel parametric array loudspeaker array, and demonstrated that compared with the delay-and-sum (DAS) algorithm, the MVDR algorithm not only effectively suppresses the side lobes but also alleviates the problem of sound blurring. Huang *et al*. [27] suggested an improved MVDR algorithm based on the enhanced virtual aperture (EVA) array to address the problem that it is difficult for MVDR to achieve robust high-resolution estimation under complex signal conditions (low SNR and limited snapshots). Simulations and experiments verified that this algorithm achieved sufficiently high azimuth resolution and low side lobes. By conducting a spectrum selection with the time-frequency domain transformation, this improved MVDR algorithm could effectively shorten the calculation time and reduce the computational volume.

Subspace-based algorithms, such as the Multiple Signal Classification (MUSIC) algorithm [28], the Estimation of Signal Parameters via Rotational Invariance Techniques (ESPRIT) algorithm [29], and their improved versions, require the estimation of the number of multipaths and eigenvalue decomposition (EVD) or singular value decomposition (SVD) of the covariance matrix, which usually leads to high uncertainty and computational complexity. Sheng *et al*. [30] proposed a fast defect localization method that combines a circular array and the MUSIC algorithm. Compared with the DAS imaging algorithm, the MUSIC algorithm is found to have higher angular resolution but relatively lower radial resolution, which is limited by the change in the orthogonality of the steering vector. Recently, data-driven or learning-based schemes have gained increasing popularity in DOA estimation problems [31-34]. Wang *et al*. [33] proposed a high-precision DOA estimation algorithm based on deep neural networks, which achieves satisfactory results for DOA estimation of systems with a larger number of antennas. Xiao *et al*. [34] designed a two-layer DOA estimation model that combines support vector machine (SVM) classification and random forest (RF) regression. Although this model can achieve real-time DOA estimation for single-target estimation, its accuracy is comparable to that of the MUSIC algorithm and has not been significantly improved. The reason lies in the fact that when the model extracts the upper triangular elements of the covariance matrix as features for learning, it loses the structural information and temporal features of the covariance matrix.

As the Convolutional Neural Network (CNN) is renowned for the excellent performance in image recognition and classification, it is suitable to analyze complex image patterns that are often difficult to distinguish and could accurately classify various types of defects [35]. With successful applications of the CNN in fields like agriculture [36], medicine [37], and computer vision [38], the industry has also begun to explore defect detection and classification methods based on CNNs [39-46]. He and Liu [44] proposed a four-stage general industrial defect detection framework that integrates the regression analysis and classification techniques, among which the first three stages use convolutional networks, while the fourth one for defect image classification uses the residual network, and they realized high-speed and high-precision defects detection on three datasets (including two public datasets). Arif *et al*. [45] innovatively proposed a CNN-SVM deep learning model with hybrid Bayesian optimization for real-time classification and prediction of surface roughness in machined surface images. Through the replacement of the Softmax layer in the Bayesian-optimized CNN model with an SVM classifier, they showed that the classification accuracy of this model is further improved by 5.83%. Said El-Hawwat *et al*. [46] proposed a two-stage ultrasonic detection method to effectively quantify the geometry of internal cracks in polyethylene (PE) pipes. Specifically, after they converted images of received signals into gray scale images through the continuous wavelet transform (CWT), the transfer learning strategy was employed to train two CNN models for crack image classification and prediction. Although the

CNN has been applied to some extent in the defect localization of pipe based on ultrasonic guided waves, there are still few studies on pipe defects classification, especially for image classification. Considering the error caused by the orthogonality between the signal sub-spaces and the noise sub-spaces in subspace-based algorithms and the feature learning error of DOA estimation using machine learning, in this study a two-dimensional MVDR focused beamforming method based on a uniform circular array is proposed to achieve accurate positioning and imaging of pipe defects. To underscore the advantages of the MVDR algorithm in this study, comparisons are performed against the CBF algorithm and the MUSIC algorithm. Meanwhile, the potential limitations of the MVDR algorithm in practical applications are also discussed. Considering the powerful feature extraction and classification performance of the CNN, this paper constructs a CNN-based model to classify the collected defect images, and evaluates the generalization ability and classification performance metrics of the proposed model. Furthermore, a comparison with other neural network models (such as AlexNet, DarkNet19, and VGG16) is conducted to demonstrate the applicability of the proposed model.

## 2. Methodologies

This paper proposes a novel MVDR-CNN-based framework for the imaging, localization, and classification of pipe defects, which integrates the near-field MVDR focused beamforming technique and CNN. As illustrated in **Fig. 1**, the overall process includes the following four steps: (1) Establish a uniform circular sensor array to collect guided wave signals from healthy and defective pipes, respectively. (2) Calculate the wave velocity using the wavelet transform fitting method. (3) Subtract the healthy baseline signal from the defective signal to obtain the defect-scattered signal. Use the MVDR algorithm to localize hole and crack defects in the pipe. Then collect magnified images of these defects as the dataset for the subsequent defect classification network. (4) Construct the CNN model, and apply Bayesian optimization to determine its hyperparameters. Train the CNN model using $k$-fold cross-validation, test the trained model with the test dataset, and assess the model's performance through various metrics.

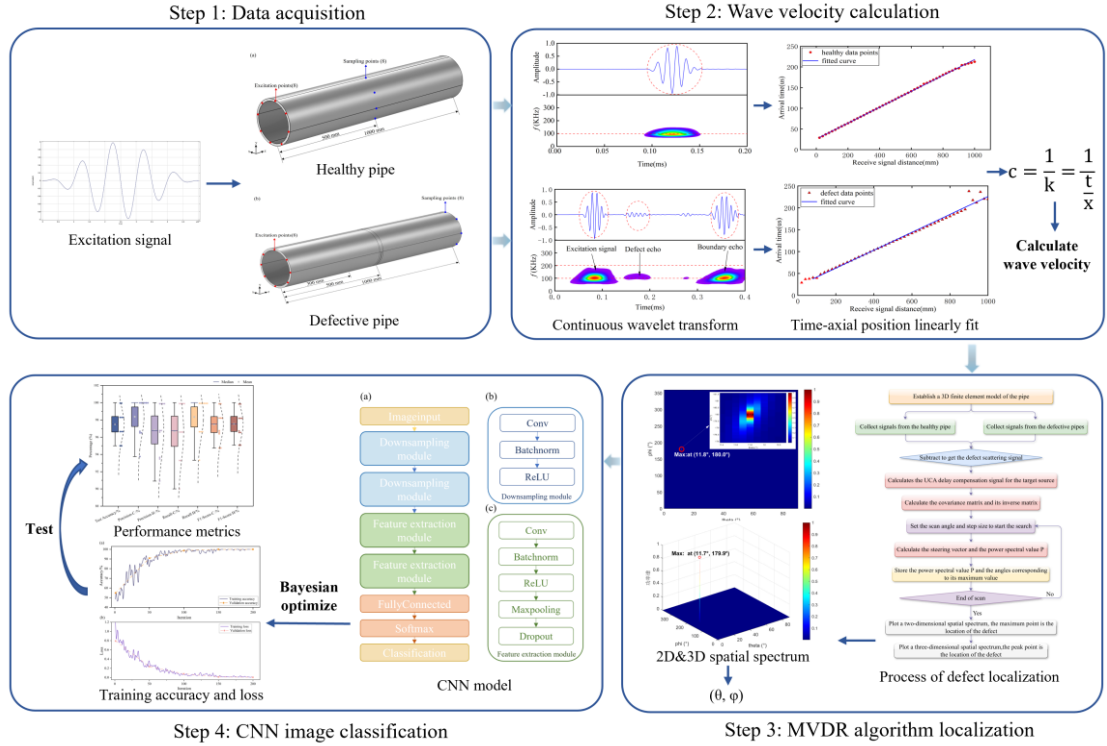

**Fig. 1**. Flowchart of defect localization and classification based on the MVDR-CNN Model.

**2.1 Selection of the excitation frequency and mode**

Ultrasonic guided waves propagating in pipes have dispersion and multimode characteristics. Usually, a frequency with fewer modes is selected as the excitation frequency to reduce the difficulty of signal analysis. According to the propagation direction, guided waves can be divided into two major categories, *i.e.*, the longitudinal and circumferential waves. Circumferential guided waves are typically employed for defect detection in the thickness direction, whereas longitudinal guided waves are commonly utilized for rapid scanning of long-distance pipelines. In the present study, the longitudinal guided wave is selected as the research object. Via solving the guided wave equation, dispersion curves are shown in **Fig. 2** in this study. The outer diameter of the pipe is 200 mm and the wall thickness is 10 mm. It is found from the group velocity curve that the number of modes increases in the frequency range above 200 kHz, and the group velocity fluctuates significantly in the range below 50 kHz. Meanwhile, compared with the L(0,1) mode, the L(0,2) mode always maintains a relatively high propagation speed and its near non-dispersive nature over the low-frequency range [47]; it could provide accurate and efficient results in the ultrasonic guided wave detection of long-distance pipes. Therefore, 100 kHz is selected as the excitation frequency in this paper, and the L(0,2) mode is adopted for subsequent researches.

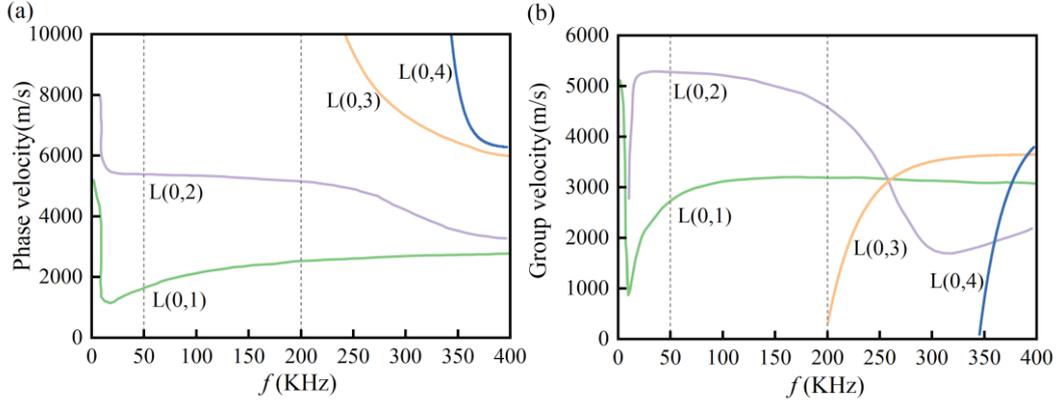

**Fig. 2**. Dispersion curves of longitudinal modes in the pipe: (a) phase velocity; (b) group velocity.

**2.2 Near-field two-dimensional MVDR focused beamforming localization**

According to the commonly used criterion [48], the propagation process is divided into near-field and far-field propagation with the cutoff distance $r = \frac{2D^2}{\lambda}$, where $r$ is the distance from the signal source to the array center, $D$ is the array aperture, and $\lambda$ is the wavelength of the signal source. In this paper, D is 200 mm, λ is the ratio of wave speed to excitation frequency, with a value of 52 mm. Thus, the value of r is 1538.5 mm. The conventional beamforming technique is based on the far-field plane wave assumption, and its time delay difference compensation amount is only related to the azimuth. In contrast, the focused beamforming technique is based on the near-field spherical wave assumption, and the spherical wave compensation for the time delay difference is carried out according to both the azimuth and elevation angles in the estimation regime; thus, this technique could access the beam location at different spatial positions.

Considering the propagation of ultrasonic guided waves in a defective pipe, this paper constructs a near-field circular sensor array model to realize the defect localization. The guided wave excited by the transducer outside the circular detection array scatters when encountering defects during the propagation process in the pipe, and the scattered signal is then received by the element sensors of the circular array. As shown in **Fig. 3**, $M$ receiving sensors are evenly distributed along the circumference with a radius of $R$ on the upper and lower surfaces of the pipe, while the defect is regarded as the sound source.

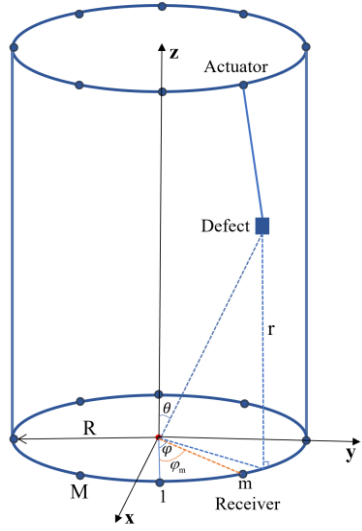

Fig. 3. Schematics of the uniform distribution of circular arrays on the pipe surface.

Taking the element 1 as the reference sensor, the output signal of the element sensor $m$ is expressed as [49]:

$$x_m(t) = e^{-j\omega\tau_m} s_m(t) + n_m(t) \quad (1)$$

where, $s_m(t)$ is the signal received by the element $m$, $n_m(t)$ is the noise signal, and $\tau_m$ is the signal delay time of the element,

$$\tau_m(\theta, \varphi) = \frac{R}{c} \sin\theta \left[\cos(\varphi_m - \varphi)\right] \quad (2)$$

where, $\theta$ is the pitch angle of the defect-scattered signal relative to the center of the circle, $\varphi$ is the azimuth angle of the projection of the defect-scattered signal on the $xOy$-plane relative to the line from the element 1 to the center of the circle, $\varphi_m = \frac{m-1}{M} 2\pi \ (m = 1, 2, \cdots, M)$ is the phase angle of each element relative to the element 1.

By performing delay compensation on the signals received by the array and superimposing the compensated signals, the output signal of the uniform circular array is written as:

$$X(t) = A(\theta, \varphi)S(t) + N(t) \quad (3)$$

where

$$X(t) = [x_1(t), x_2(t), \cdots, x_M(t)] \quad (4)$$

$$A(\theta, \varphi) = \left[1, e^{-j\omega\tau_1}, e^{-j\omega\tau_2}, \cdots, e^{-j\omega\tau_M}\right]^T \quad (5)$$

$$S(t) = [s_1(t), s_2(t), \cdots, s_M(t)] \quad (6)$$

$$N(t) = [n_1(t), n_2(t), \cdots, n_M(t)] \tag{7}$$

$$a_m(\theta, \varphi) = e^{-j\omega\tau_m(\theta,\varphi)}, m = 1, 2, \cdots, M \tag{8}$$

The superimposed output can be expressed as:

$$Y(t) = W^H(\theta, \varphi) X(t) \tag{9}$$

where $W$ is the weighting vector. The output power is written as:

$$P(\theta, \varphi) = W^H(\theta, \varphi) R W(\theta, \varphi) \tag{10}$$

where $R = E[X(t)X(t)^H]$ is the covariance matrix.

The objective of the MVDR focused beamforming algorithm is to maintain the power output of the desired signal and minimize the power output of interferences and noises, *i.e.*, to solve the following constrained minimization problem:

$$\min \ W^H(\theta, \varphi) R W(\theta, \varphi) \ \text{s.t.} \ W^H(\theta, \varphi) a(\theta, \varphi) = 1 \tag{11}$$

The optimal weighting vector obtained by solving the above equation is,

$$W = \frac{R^{-1} a(\theta, \varphi)}{a^H(\theta, \varphi) R^{-1} a(\theta, \varphi)} \tag{12}$$

The spatial spectrum function of the MVDR focused beamforming is,

$$P(\theta, \varphi) = W^H(\theta, \varphi) R W(\theta, \varphi) = \frac{1}{a^H(\theta, \varphi) R^{-1} a(\theta, \varphi)} \tag{13}$$

Since the actually received data matrix has a finite length, the maximum likelihood estimation $\hat{R}$ is used to replace $R$, where $\hat{R}$ is given by,

$$\hat{R} = \frac{XX^H}{N} \tag{14}$$

where, $N$ is the number of snapshots. Therefore, the output power is expressed as,

$$P(\theta, \varphi) = W^H(\theta, \varphi) R W(\theta, \varphi) = \frac{1}{a^H(\theta, \varphi) \hat{R}^{-1} a(\theta, \varphi)} \tag{15}$$

For this method, defect location identification is realized by constructing the spatial spectrum function. This method scans two angular dimensions (i.e., pitch angle $\theta$ and azimuth angle $\varphi$) in the detection region point by point and calculates the spatial spectrum response at each scanning point. When the scanning point coincides with the actual defect location, the spatial spectrum function reaches its theoretical maximum. The position corresponding to this peak point is identified as the defect location. **Fig. 4** shows the flowchart of the MVDR-based defect localization algorithm.

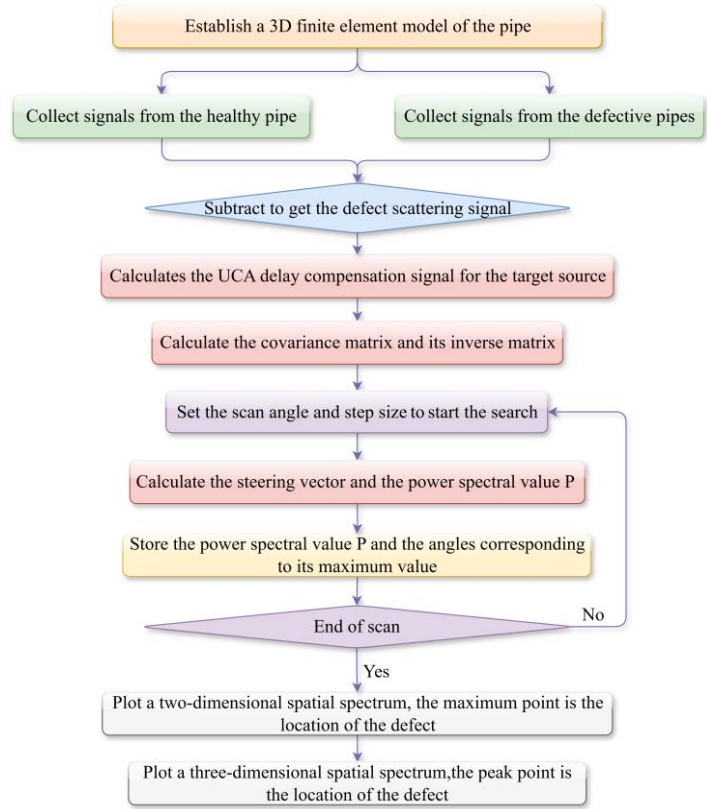

**Fig. 4**. Flowchart of near-field two-dimensional defect localization imaging based on MVDR focused beamforming.

## 3. Results and Discussion

### 3.1 Wavelet transform fitting and identification

Based on the finite element (FE) analysis method, this paper constructs a three-dimensional pipe model to simulate the propagation characteristics of ultrasonic guided waves. The geometric parameters of the model are set as follows: the outer diameter 200 mm, wall thickness 10 mm, and axial length 1000 mm. As shown in **Fig. 5**, the left surface of the model is set as the excitation end. Once the guided wave signal is excited, the guided wave propagates from the left end of the pipe to the right end. After being reflected by the right end surface, it propagates back to the left end surface again. The material parameters used for FE analysis are detailed in **Table 1**.

**Table 1**. Material parameters of the pipe model.

| material | $\rho$ (g/cm³) | $E$ (GPa) | $v$ |
|---|---|---|---|
| X70 steel | 7.93 | 211 | 0.29 |

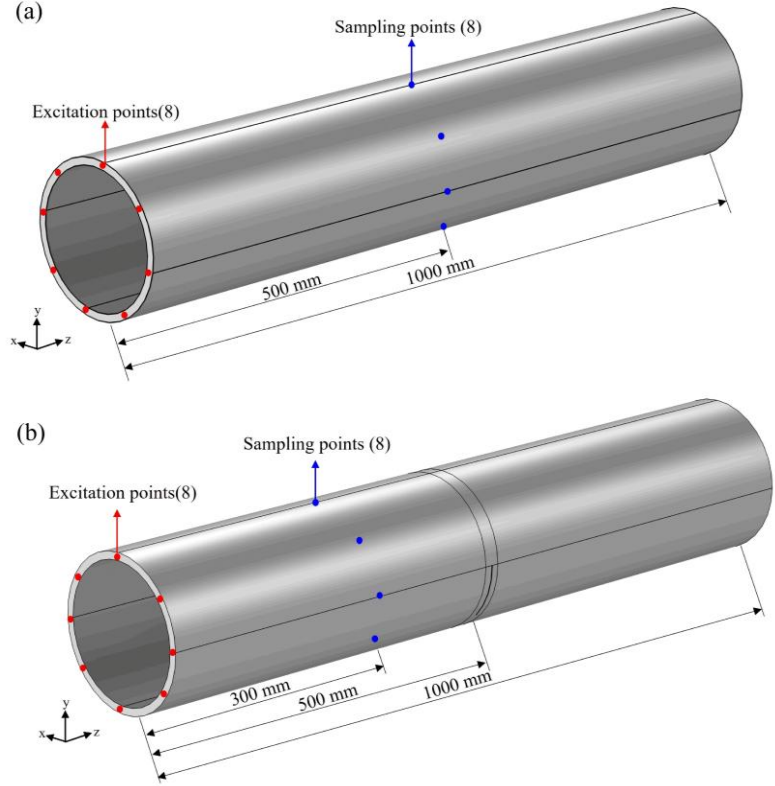

**Fig. 5**. Three-dimensional finite element model of (a) healthy pipe, (b) defective pipe.

A defective model is also established on the basis of the healthy pipe model. A crack defect with the radial depth of 5 mm and circumferential dimension angle of 45 ° is set at 500 mm along the axial direction, as shown in **Fig. 5**b. To obtain the defect response signal, eight signal sampling points are evenly arranged along the pipe circumference at an axial distance of 300 mm from the excitation end. Superimpose the received signals and perform averaging processing. The chosen excitation signal is a Hanning windowed five-cycle sine wave with a center frequency of 100 kHz. The mathematical expression of the excitation signal is:

$$S(t) = 0.5 \times \left(1 - \cos\left(\frac{2\pi f_c t}{n}\right)\right) \cdot \sin(2\pi f_c t) \tag{16}$$

where $t$ is the time step, $n$ is the number of cycles of the Hanning window, $f_c$ is the excitation center frequency.

As shown in **Fig. 6**, time-frequency analysis is performed on the excitation signal of the healthy pipe and the guided wave propagation signal of the defective pipe via the CWT. In **Fig. 6**a, the main frequency components of the excitation signal are stably concentrated at about 100 kHz, which is consistent with the preset excitation frequency. Along the time axis, these main frequency components are mainly concentrated in the range of 0.1 ~ 0.14 ms and show a clear time-domain distribution. This indicates that no significant dispersion occurs during the propagation of the excitation signal, and it also verifies that in a healthy pipeline, the propagation process of the guided wave signal is stable without significant scattering and reflection. In **Fig. 6**b, when the guided wave propagates in the defective pipe, the main frequency of the excitation signal remains stable at approximately 100 kHz. A defect echo signal appears around 0.17 ms on the time axis, and its main frequency component is slightly higher than 100 kHz. This frequency shift phenomenon is mainly caused by the influence of the local structure during the defect reflection process and the frequency coupling effect. At about 0.35 ms, an end surface echo appears as the guided wave propagates to the end surface. Its main frequency is close to the excitation frequency, but the energy is slightly attenuated compared with both the excitation signal and the defect echo. In addition, it can also be observed from the time-domain diagram (**Fig. 6**b) that compared with other modes, the L(0,2) mode exhibits less attenuation during propagation, which is conducive to long-distance defect detection..

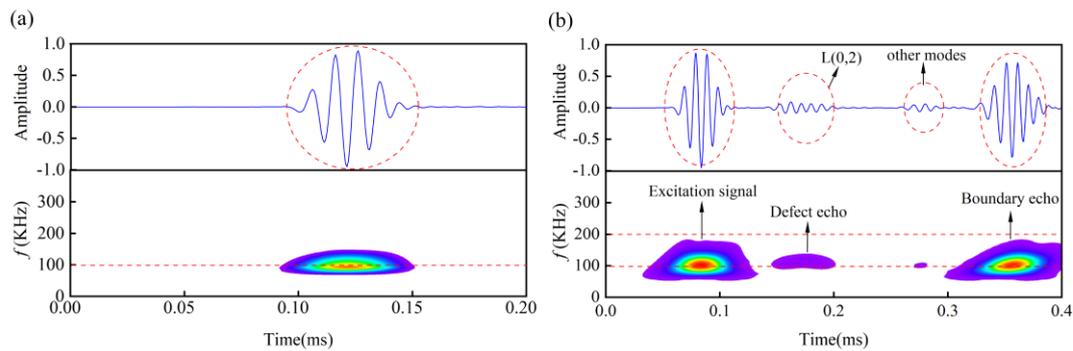

**Fig. 6**. Time-frequency analysis via CWT of (a) excitation signal in the healthy pipe, (b) guided wave propagation signal in the defective pipe.

The wave velocity is measured both for models of the healthy and defective pipes. Specifically, 50 sets of time-domain signals are collected for each model within the range of 20 mm to 1000 mm along the axial direction of the pipe. The time information corresponding to the peak points of the 100 kHz main frequency component is extracted via the CWT, and the obtained 'time-axial position' dataset is linearly fitted. In **Fig. 7**a, the slope of the straight line obtained through linear fitting is 0.1901 μs/mm, so the propagation velocity of the guided wave in a healthy pipe is computed as 1/(0.1901μs/mm) ≈ 5260 m/s. In **Fig. 7**b, for the defective pipe, the slope of the straight line obtained through linear fitting is 0.1923 μs/mm and the guided wave velocity is calculated as 1/(0.1923 μs/mm) ≈ 5200 m/s. Although there exist a few data points at both ends deviating from the linearity, the results agree well with the fitting line, indicating that the velocity of the guided wave remains relatively stable during the propagation process.

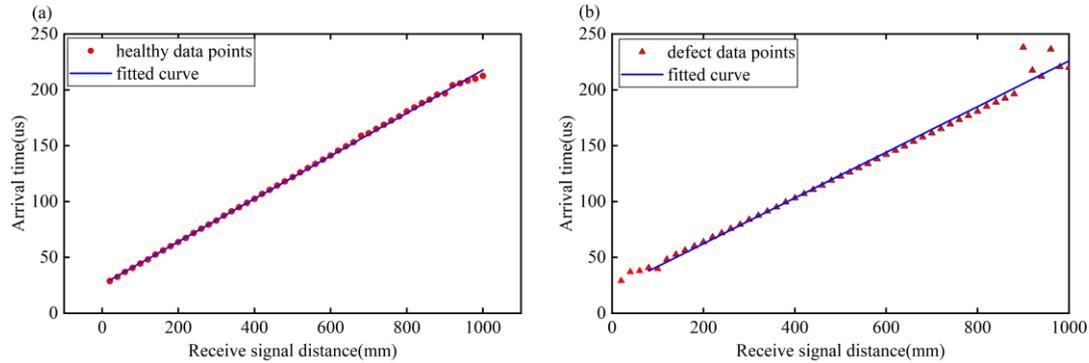

Fig. 7. Fitting of the axial position-propagation time data of the guided wave in the (a) healthy pipe, (b) defective pipe.

Different from the present wavelet transform fitting method, the classical pulse-echo method to measure the guided wave velocity works as follows: 1) After the excitation sensor generates the guided waves, the guided waves propagate along the axial direction of the pipe; 2) The guided waves are received by the receiving sensors, then continue to propagate to the end-face of the pipe and be reflected; 3) The reflected waves are received by the receiving sensors again. Considering the above propagation process, where the propagation time of the guided wave is $t$, the propagation distance is $l$ (twice the distance between the sampling point and the pipe end-face), and the wave velocity is $c$, the simple correlation can be established as $l = ct$. The positions of the signal sampling points for measuring the echo signal are shown in **Fig. 5** above, and the corresponding measurement results are presented in **Fig. 8**. As shown in **Fig. 8**, by extracting the envelope of the time-domain signal of the guided wave propagation, the distance $l_1$ propagated by $t_1$ is 1000 mm, and the distance $l_2$ propagated by $t_2$ is 1400 mm. Thus $v_1$ is 5128 m/s, and $v_2$ is 5037 m/s. Clearly, compared with the CWT fitting method, the pulse-echo method shows greater deviation between the measured wave velocity and the theoretical value.

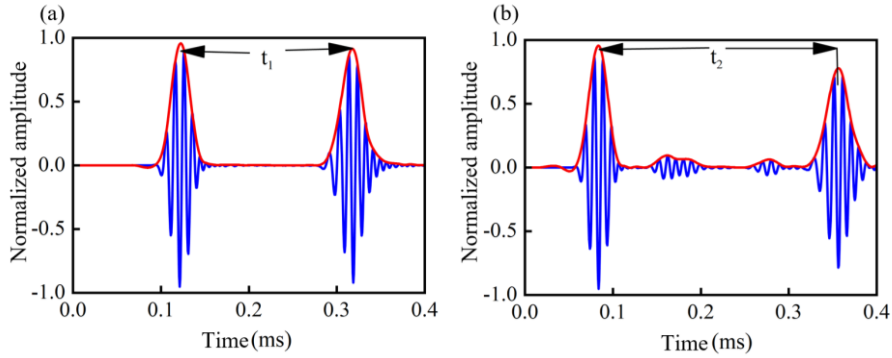

**Fig. 8**. Envelope diagrams of guided wave response signals in (a) healthy pipe; (b) defective pipe.

**Table 2** compares these two methods for calculating the velocity of the L(0,2) guided wave. Notably, for the healthy and defective pipes respectively, the CWT fitting method has a lower wave velocity relative error than the pulse-echo method by 2.48% and 3.06%. This result indicates the superiority of the wavelet transform signal processing method. In addition, due to the existence of defects, the wave velocity measurement error of the defective pipe is larger than that of the healthy pipe, especially for the pulse-echo method, indicating that the existence of defects does indeed affect the propagation of the guided waves.

**Table 2**. Comparison of velocity calculation and relative error of the guided wave L(0,2) mode in pipes (theoretical value: 5325 m/s).

| Method | Pipe Type | Calc. Value (m/s) | Relative Error |
| --- | --- | --- | --- |
| CWT fitting method | Healthy | 5260 | 1.22% |
| | Defective | 5200 | 2.35% |
| pulse-echo method | Healthy | 5128 | 3.70% |
| | Defective | 5037 | 5.41% |

**3.2 Defect imaging via near-field MVDR focused beamforming**

As shown in **Fig. 9**, eight piezoelectric elements are evenly arranged circumferentially at one end of the pipe to form the excitation array for signal excitation, and a circular receiving array—also consisting of eight piezoelectric elements—is correspondingly arranged at the other end of the pipe. For the excitation process, each actuation step drives one piezoelectric element individually; eight sequential excitations are performed to activate the actuators E1 to E8 one by one. The pipe is then unfolded along its straight generatrix into a two-dimensional rectangle (*i.e.*, the pipe length × pipe circumference). The coordinate positions of each piezoelectric element in the unfolded plane are listed in **Table 3**. The area marked by the red box in **Fig. 9** corresponds to the position of the defect after unfolding.

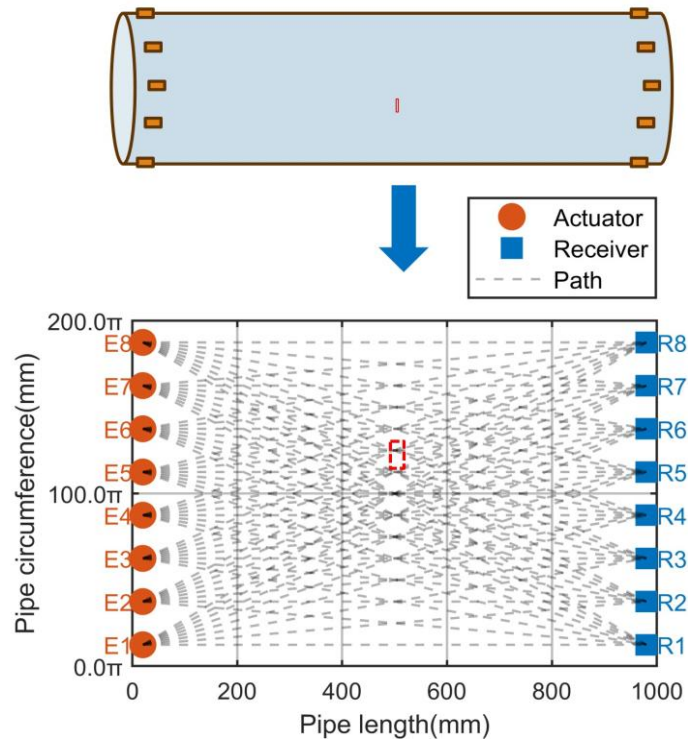

**Fig. 9**. Schematic of actuator/receiver distribution and guided wave propagation paths.

Table 3. Coordinates of 16 piezoelectric sensing elements.

| Actuator | Coordinates (mm) | Receiver | Coordinates (mm) |
|---|---|---|---|
| E1 | $(20, 12.5\pi)$ | R1 | $(980, 12.5\pi)$ |
| E2 | $(20, 37.5\pi)$ | R2 | $(980, 37.5\pi)$ |
| E3 | $(20, 62.5\pi)$ | R3 | $(980, 62.5\pi)$ |
| E4 | $(20, 87.5\pi)$ | R4 | $(980, 87.5\pi)$ |
| E5 | $(20, 112.5\pi)$ | R5 | $(980, 112.5\pi)$ |
| E6 | $(20, 137.5\pi)$ | R6 | $(980, 137.5\pi)$ |
| E7 | $(20, 162.5\pi)$ | R7 | $(980, 162.5\pi)$ |
| E8 | $(20, 187.5\pi)$ | R8 | $(980, 187.5\pi)$ |

As shown in **Fig. 9**, when one actuator emits an excitation signal, the receivers at the right end can receive eight defect-scattered signals. One excitation is applied to one actuator. Since one excitation is applied to one actuator, eight healthy baseline signals can correspondingly be obtained in the healthy pipe model. The defect-scattered signal for the corresponding actuator is obtained by subtracting the healthy baseline signal from the defective signal. After filtering this defect-scattered signal and inputting it into the MVDR algorithm, a single defect location can be derived. Sequential switching through the eight excitation channels allows the acquisition of eight sets of defect location results. The core positioning parameters of this algorithm are the pitch angle $\theta$ (of the defect-scattered signal relative to the center of the circle) and the azimuth angle $\varphi$ (of the defect-scattered signal's projection on the *xOy*-plane relative to the line from the array element 1 to the circle center). The pitch angle $\theta$ is defined as follows:

$$\theta = \arctan\frac{R}{r} \qquad (21)$$

where, $R$ denotes the outer diameter of the pipe, and $r$ represents the axial distance from the defect to the circular array.

As shown in **Fig. 9**, a hole defect is positioned 500 mm along the axial direction of the pipe, with a radius of 12 mm. Taking sensor R1 as the reference sensor, the hole defect is located at (11.77°, 180.00°) in the ($\theta$, $\varphi$) diagram. Using the MVDR algorithm described in **Section 2.2**, the normalized spatial spectrum of the hole defect can be obtained. Without loss of generality, **Fig. 10** presents the imaging and localization results when excitation is provided by actuator E5. The region marked by the red circle is the defect position. The localization results and relative errors of all actuators are summarized in **Table 4**.

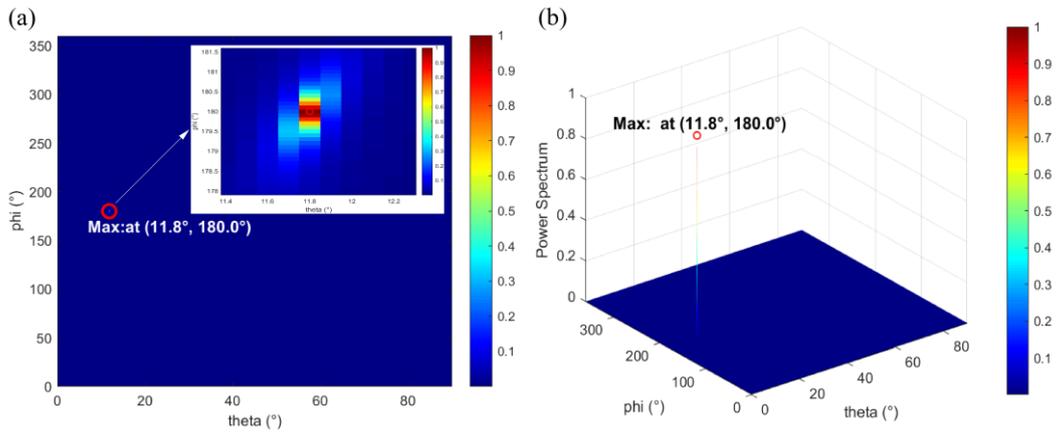

Fig. 10. MVDR imaging spatial spectra of the hole defect: (a) Two-dimensional spatial spectrum; (b) Three-dimensional spatial spectrum.

Table 4. Localization results of hole defect excited by different actuators.

| Actuator | $\theta/°$ | $\varphi/°$ | $r$/mm | $\delta_L$ | $\delta_C$ |
|---|---|---|---|---|---|
| E1 | 11.8 | 180.1 | 478.7 | 0.27% | 0.05% |
| E2 | 11.8 | 179.8 | 478.7 | 0.27% | 0.10% |
| E3 | 11.8 | 180.2 | 478.7 | 0.27% | 0.10% |
| E4 | 11.8 | 179.9 | 478.7 | 0.27% | 0.05% |
| E5 | 11.8 | 180.0 | 478.7 | 0.27% | 0 |
| E6 | 11.8 | 179.9 | 478.7 | 0.27% | 0.05% |
| E7 | 11.7 | 179.8 | 482.8 | 0.58% | 0.10% |
| E8 | 11.8 | 180.3 | 478.7 | 0.27% | 0.15% |

Similarly, a crack defect with a circumferential angle of 30° is positioned 500 mm along the axial direction of the pipe, and is located at (11.77°, 202.50°) in the ($\theta$, $\varphi$) diagram. The spatial spectrum of the crack defect excited by actuator E5 is shown in **Fig. 11**. The localization results of the crack defect for all actuators are summarized in **Table 5**.

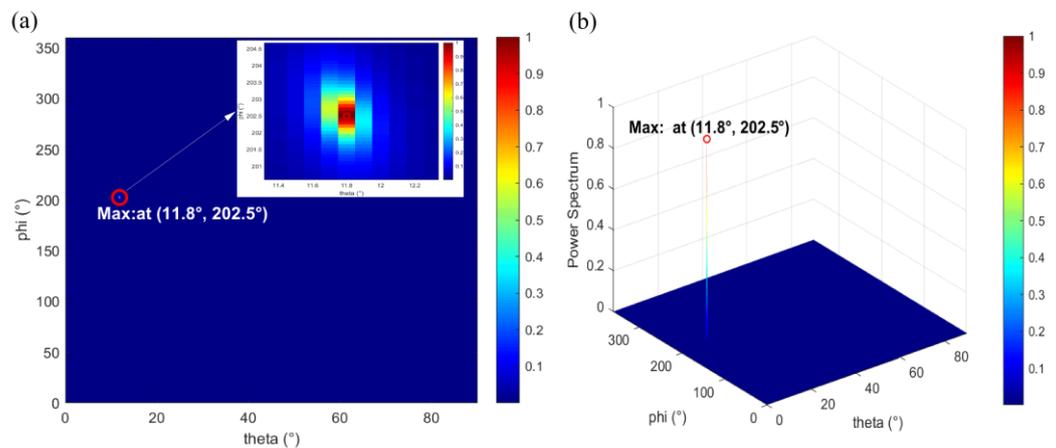

Fig. 11. MVDR imaging spatial spectra of the crack: (a) Two-dimensional spatial spectrum; (b) Three-dimensional spatial spectrum.

Table 5. Localization results of crack defect excited by different actuators.

| Actuator | $\theta$/° | $\varphi$/° | $r$/mm | $\delta_L$ | $\delta_C$ |
|---|---|---|---|---|---|
| E1 | 11.8 | 202.7 | 478.7 | 0.27% | 0.09% |
| E2 | 11.8 | 202.3 | 478.7 | 0.27% | 0.09% |
| E3 | 11.8 | 202.8 | 478.7 | 0.27% | 0.14% |
| E4 | 11.8 | 202.4 | 478.7 | 0.27% | 0.05% |
| E5 | 11.8 | 202.5 | 478.7 | 0.27% | 0 |
| E6 | 11.7 | 202.5 | 482.8 | 0.58% | 0 |
| E7 | 11.8 | 202.4 | 478.7 | 0.27% | 0.05% |
| E8 | 11.8 | 202.1 | 478.7 | 0.27% | 0.19% |

The simulation results presented in **Fig. 10 & 11** indicate that the MVDR focused beamforming algorithm exhibits high locating accuracy for both hole and crack defects. In the two-dimensional localized magnified image, the power spectrum values at the defect locations are significantly higher than those of the surrounding non-defective areas. The point with the maximum power spectrum marked by a red circle, corresponds to the defect location. Meanwhile, this location corresponds to the only extreme point in the spatial spectrum, with no other artifacts. Furthermore, the three-dimensional spatial spectrum exhibits a sharp single-peak characteristic with no other side lobes. This thus demonstrates that the algorithm possesses excellent target directivity, as well as high resolution in both pitch and azimuth angles.

In **Table 4 & 5**, by averaging the relative errors of eight locating tests, the average longitudinal errors $\delta_L$ of the hole and crack defect are 0.14% and 0.075%, respectively, and the average circumferential errors $\delta_C$ are 0.14% and 0.076%, respectively. Analysis indicates that the circumferential localization accuracy of the defects is higher than the longitudinal localization accuracy. In addition, the longitudinal errors are generally larger than circumferential errors. The main reason for this is that the longitudinal distance of the defect needs to be calculated using the tangent function, which exhibits nonlinear and rapid growth characteristics within the range of 0°-90°, thereby introducing errors in the axial localization of the defect.

To highlight the advantages of the MVDR algorithm in this study, the crack defect mentioned above is taken as an example, and the localization results obtained via the CBF algorithm and the MUSIC algorithm are shown in **Fig. 12**. All results shown in the figure are excited by actuator E5. As can be seen from the figure, the localization results of the three algorithms are all relatively accurate. However, the resolution of the CBF algorithm is significantly lower than that of the MVDR and MUSIC algorithm.

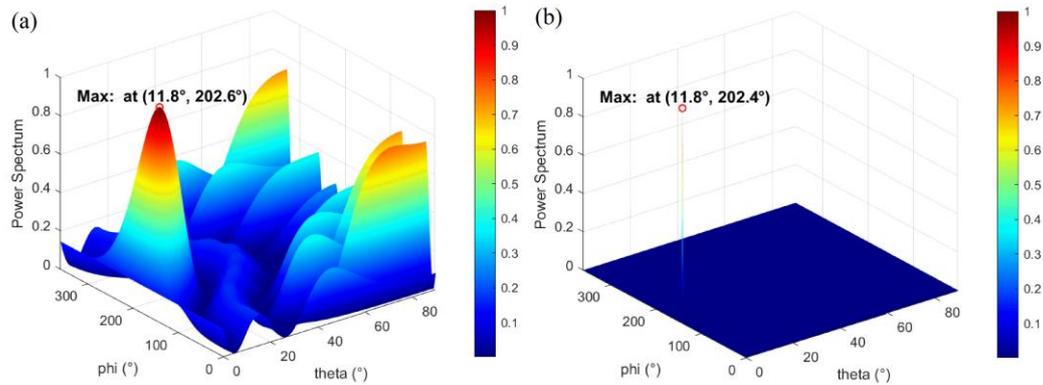

**Fig. 12**. Spatial spectra of different algorithms for crack defect localization:(a) CBF algorithm;(b) MUSIC algorithm.

Given the significantly prolonged execution time caused by the extensive computational demands of the MUSIC algorithm, we further evaluated the computational efficiency of three algorithms (MVDR, CBF, and MUSIC) through numerical experiments. As shown in **Table 6**, when the scanning precision is set to 0.1, the average imaging time of the MVDR algorithm—for 900 pitch angle indexings and 3600 azimuth angle indexings—is 55.56 s, which is the shortest time among the three algorithms. In terms of computational complexity, the runtime of the MVDR algorithm falls within a reasonable range; thus, it is industrially feasible regarding computational efficiency.

Table 6. Localization time of the crack defect for different algorithms.

| Actuator | Time (s) | | |
| --- | --- | --- | --- |
| | CBF | MVDR | MUSIC |
| E1 | 56.51 | 56.14 | 63.76 |
| E2 | 56.05 | 55.28 | 62.09 |
| E3 | 55.92 | 55.50 | 62.10 |
| E4 | 58.08 | 55.66 | 65.13 |
| E5 | 55.88 | 55.38 | 61.67 |
| E6 | 55.96 | 55.71 | 62.27 |
| E7 | 55.73 | 55.48 | 62.34 |
| E8 | 55.92 | 55.33 | 62.33 |
| Average | 56.26 | 55.56 | 62.71 |

SNR is defined as the ratio of signal power to noise power in a system. After calculating the power of the input signal in the algorithm, Gaussian white noise with the corresponding SNR is added to simulate a real-world noise environment. To explore the influence of input signal SNR on defect localization accuracy, the hole defect (with a radius of 12 mm) is taken as an example. The SNR range is set from -10 dB to 35 dB at an interval of 5 dB, and actuator E5 is used for defect localization, without loss of generality. As shown in **Fig. 13**a, as the SNR increases, the pitch angle $\theta$, azimuth angle $\varphi$, and axial distance $r$ gradually approach their target values. For low SNRs, **Fig. 13**b further shows that the localization error deceases gradually with increasing SNR. When the SNR reaches 30 dB, the relative error reaches its minimum. In fact, even at an SNR of 5 dB, both the longitudinal and circumferential errors are already below 1%. This proves that the MVDR algorithm still maintains high localization resolution under low-SNR conditions.

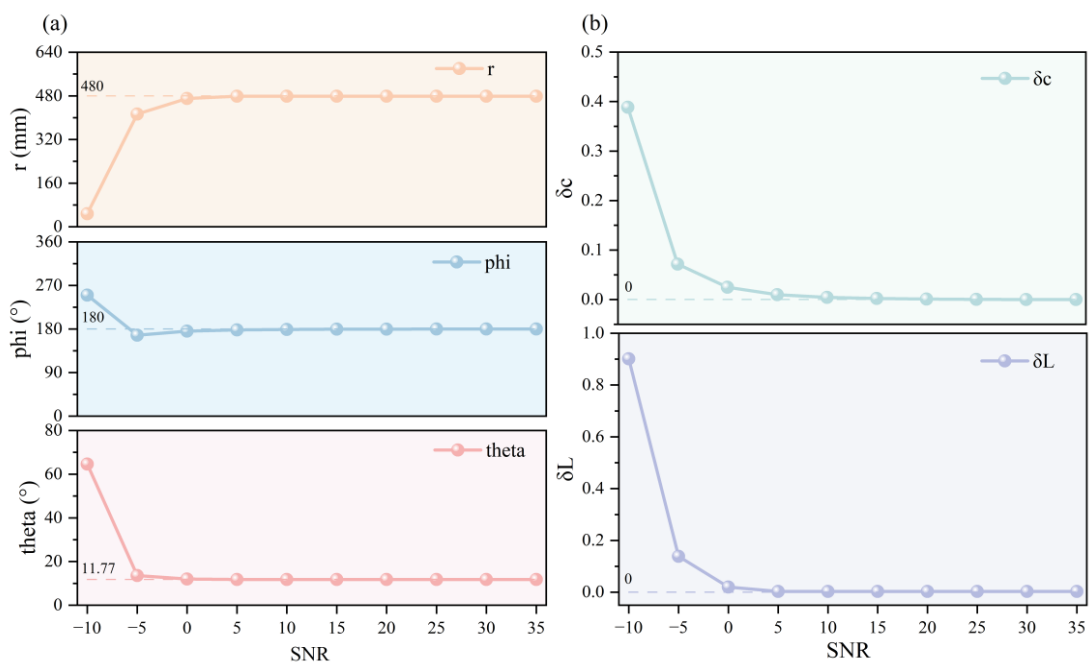

**Fig. 13**. Localization errors of the hole defect under different SNRs (baseline: $r$ = 480 mm, $\varphi$ = 180°, $\theta$ = 11.77°). (a) longitudinal distance $r$, azimuth angle $\varphi$, and pitch angle $\theta$; (b) circumferential error $\delta_C$ and longitudinal error $\delta_L$.

**3.3 CNN-driven image classification analysis**

By observing the locally magnified regions of the two-dimensional spatial spectrograms in **Fig. 10**a

& **11**a, it is found that the spectrogram image of the hole and crack defect are highly similar. Thus, it is difficult to distinguish between them by visual observation alone. This highlights the necessity of introducing an automated feature extraction method. Therefore, this paper proposes a CNN-based framework for feature extraction and defect classification.

**3.3.1 Dataset preprocessing and partitioning**

Data augmentation techniques such as image rotation, horizontal/vertical flipping, scaling, and brightness adjustment were adopted [50]. These preprocessing operations not only expand the dataset size but also improve the model's generalization ability for unknown samples. Furthermore, they effectively suppress the risk of overfitting and enhance the model's robustness[51]. Based on the magnified views of two-dimensional spatial spectrograms of the hole defect and crack defect obtained via the MVDR algorithm, a high-quality defect classification dataset was constructed as follows: first, the aspect ratio and color mapping range of the spectrogram images were uniformly adjusted to ensure the visual clarity of defect-related features; then, after performing the aforementioned image enhancement operations, the final defect classification dataset was established. To ensure an adequate volume of training data and an independent test set, the dataset was randomly partitioned. Specifically, 15% of the dataset was allocated as the test set, while the remaining 85% was used for the training and validation sets. Notably, the test set is only used to evaluate the classification accuracy after network training is completed and does not participate in the training or validation processes of the network model.

To enhance the robustness of the training results and the reliability of the generalization ability evaluation, the $k$-fold cross-validation technique was employed for the model's training and validation processes. In principle, $k$-fold cross-validation [52] uniformly and randomly divides the original dataset into $k$ non-overlapping subsets (also referred to as "folds"). For the $i$-th iteration (where $i = 1, 2, \cdots, k$), the $i$-th fold is designated as the validation set, and the remaining $k$-1 folds serve as the training set. This cross-validation is repeated $k$ times, and the average of the $k$ iteration results is taken as the final generalization error $E$ of the model [53, 54], as illustrated in **Fig. 14**. Through multiple iterations and averaging, $k$-fold cross-validation can effectively avoid the contingency caused by a single division of the training and the validation set. Given the small sample size and the need to avoid underfitting, the data sets (excluding the test set) were divided into 10 subsets using *k*-fold cross-validation (*i.e.*, *k* = 10) [55, 56].

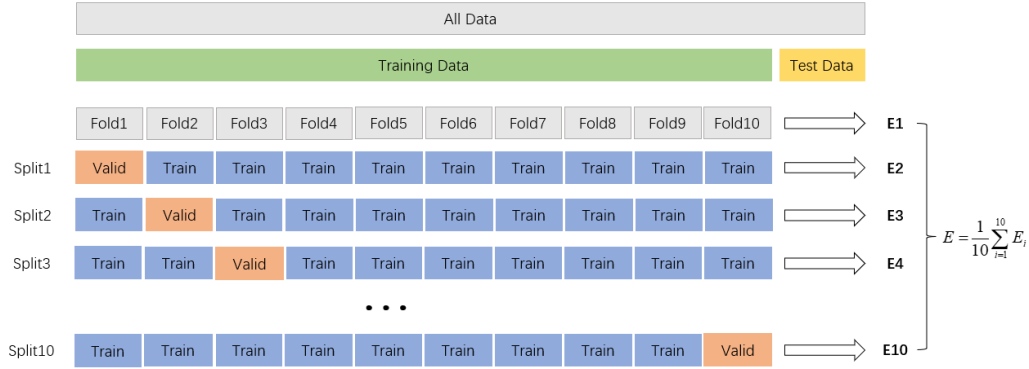

**Fig. 14**. Schematic diagram of the *k*-fold cross-validation (*k* = 10).

**3.3.2 Design of the CNN architecture**

As presented in **Fig. 15**, a 20-layer deep convolutional neural network (CNN) is designed, and the detailed parameters are listed in **Table 7**. This network adopts a progressive feature extraction strategy and achieves efficient feature representation and pattern recognition through a three-stage structure, which consists of a downsampling module, a feature learning module, and a classification module.

For the three-channel training images with an input size of 567×557 pixels, their large dimensions necessitate size compression. Thus, two downsampling modules are used, with convolutional layers employed instead of pooling layers for this purpose. Compared with traditional pooling layers, convolutional downsampling is less prone to losing image features while effectively reducing image size. The number of convolution kernels in the first layer is determined through experiments. Among numbers of 16,32, 64, 128, and 256, 128 yielded the highest classification accuracy. For the number of convolution kernels in the subsequent layers, a gradual decreasing approach following powers of 2 is adopted, primarily to cut the model's parameter count. Modern computing hardware processes such-sized data more efficiently, which supports this choice. After each convolutional layer, a batch normalization layer is applied to normalize the data, and a ReLU activation function layer is added to introduce non-linearity into the data.

After two downsampling modules, the image size is reduced to 142×140 pixels. A two-level deep convolutional structure is then employed for further feature extraction and dimension compression. The difference is that a Maxpooling layer and a dropout layer (with a dropout rate of 0.1) are added. The dropout layer helps mitigate overfitting, while the Maxpooling layer extracts local maxima to enhance feature prominence and further reduce the image size. During the convolution process, if the image size is not an integer multiple of the convolution kernel size, zero-padding is automatically applied to the image edges. A fully connected layer is placed at the end of the network; the output of this layer is then passed to a softmax layer for probability calculation, and finally, the classification layer outputs the defect classification results.

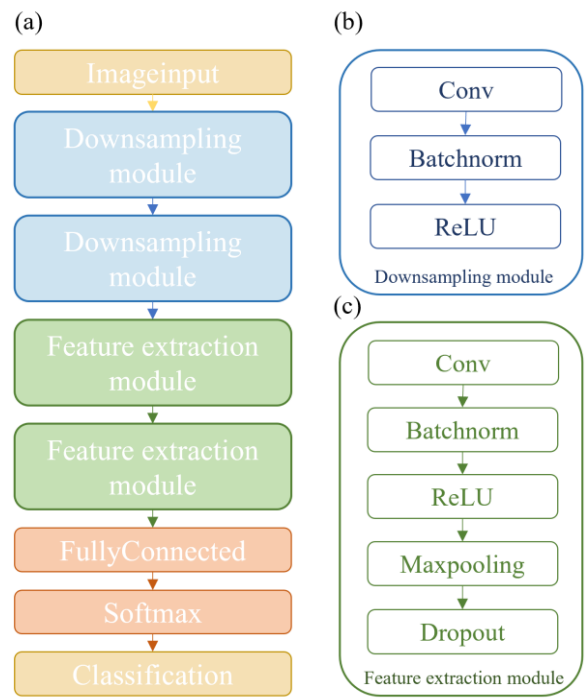

**Fig. 15.** (a) The Architecture of the proposed CNN model; (b) Downsampling module;(c) Feature extraction module.

Table 7. Detailed properties of each layer in the proposed CNN model

| Layer number | 1 | 2 | 3 | 4 | 5 | 6 | 7 |
|---|---|---|---|---|---|---|---|
| Layer type | Input | Conv1 | Bat.Norm.1 | Relu1 | Conv2 | Bat.Norm.2 | Relu2 |
| Filter size | - | 1×1 | - | - | 1×1 | - | - |
| Stride | - | 2×2 | - | - | 2×2 | - | - |
| Nodes | 3 | 128 | 128 | 128 | 64 | 64 | 64 |

| Output size | 567×557 | 284×279 | 284×279 | 284×279 | 142×140 | 142×140 | 142×140 |
|---|---|---|---|---|---|---|---|
| Layer number | 8 | 9 | 10 | 11 | 12 | 13 | 14 |
| Layer type | Conv3 | Bat.Norm.3 | Relu3 | Dropout1 | Pooling1 | Conv4 | Bat.Norm.4 |
| Filter size | 3×3 | - | - | - | 3×3 | 3×3 | - |
| Stride | 2×2 | - | - | - | 2×2 | 2×2 | - |
| Nodes | 32 | 32 | 32 | 32 | 32 | 16 | 16 |
| Output size | 71×70 | 71×70 | 71×70 | 71×70 | 36×35 | 18×18 | 18×18 |
| Layer number | 15 | 16 | 17 | 18 | 19 | 20 | |
| Layer type | Relu4 | Dropout2 | Pooling2 | Fc | Softmax | Classoutput | |
| Filter size | - | - | 3×3 | - | - | - | |
| Stride | - | - | 2×2 | - | - | - | |
| Nodes | 16 | 16 | 16 | 2 | 2 | 2 | |
| Output size | 18×18 | 18×18 | 9×9 | 1×1 | 1×1 | 1×1 | |

**3.3.3 Training of the CNN and analysis of test results**

Owing to the randomness of the input images during the cross-validation process, there exists fluctuations in the feature extraction results. Furthermore, during each training process, the weights and the learning rate change, even leading to various training results. Additionally, manually searching for the hyperparameters of the convolutional network is complex and time-consuming. Therefore, the Bayesian optimization algorithm (BOA) is employed to enable automatic hyperparameter search for the model. The core idea of BOA is to construct a surrogate model—specifically a Gaussian Process—to approximate the objective function using existing data. The optimization process is then guided by an acquisition function, which intelligently selects the next points to evaluate [57, 58]. Specifically, the hyperparameter search space, CNN architecture, and optimization objective function are inputted into the BOA. In present study, the BOA was simulated for a maximum of 30 iterations, and it selected the hyperparameter values for the CNN model that yield the best performance in classifying defect-containing images. The hyperparameter set and optimization results of the Bayesian search for the CNN model are presented in Table 8.

Table 8. Hyperparameter set from Bayesian optimization search for the CNN model

| Hyperparameters | Hyperparameter Space | Optimal Value |
|---|---|---|
| Initial Learn Rate | [0.01, 0.05] | 0.01 |
| Learn Rate Drop Period | [10, 50] | 30 |
| Learn Rate Drop Factor | [0.1, 0.9] | 0.5 |
| Validation Frequency | [10, 60] | 50 |
| Max Epochs | [20, 60] | 50 |
| Mini Batch Size | [8, 128] | 64 |
| L2Regularization | [0.001, 0.005] | 0.003 |

Using these optimized hyperparameters, the CNN model is trained for the training dataset for a total 200 iterations. Specifically, Stochastic Gradient Descent with Momentum (SGDM)—with a momentum coefficient of 0.9—is selected as the optimizer. The learning rate is initially set to 0.01 and adopts a piecewise decay schedule. After 30 epochs—where 1 epoch consists of 4 iterations (corresponding to a total of 120 iterations)—it drops to 0.005. The validation accuracy and the final loss data are presented in **Table 9**, while **Fig. 16** illustrates the validation accuracy and loss curves for the final training run of cross-validation. The average validation accuracy during training is 99.8%. The validation loss fluctuates in the range of 0.0062 to 0.0130, with an average of 0.0090, which indicates a low and stable loss level close to zero.

Table 9. Validation accuracy and final validation loss of the CNN Model.

| Number | Validation Accuracy /% | Final Validation Loss |
|---|---|---|
| 1 | 100.00 | 0.0073 |
| 2 | 100.00 | 0.0089 |
| 3 | 99.67 | 0.0097 |
| 4 | 100.00 | 0.0130 |
| 5 | 100.00 | 0.0084 |
| 6 | 100.00 | 0.0062 |
| 7 | 98.33 | 0.0098 |
| 8 | 100.00 | 0.0084 |
| 9 | 100.00 | 0.0058 |
| 10 | 100.00 | 0.0122 |
| Average | 99.80 | 0.0090 |

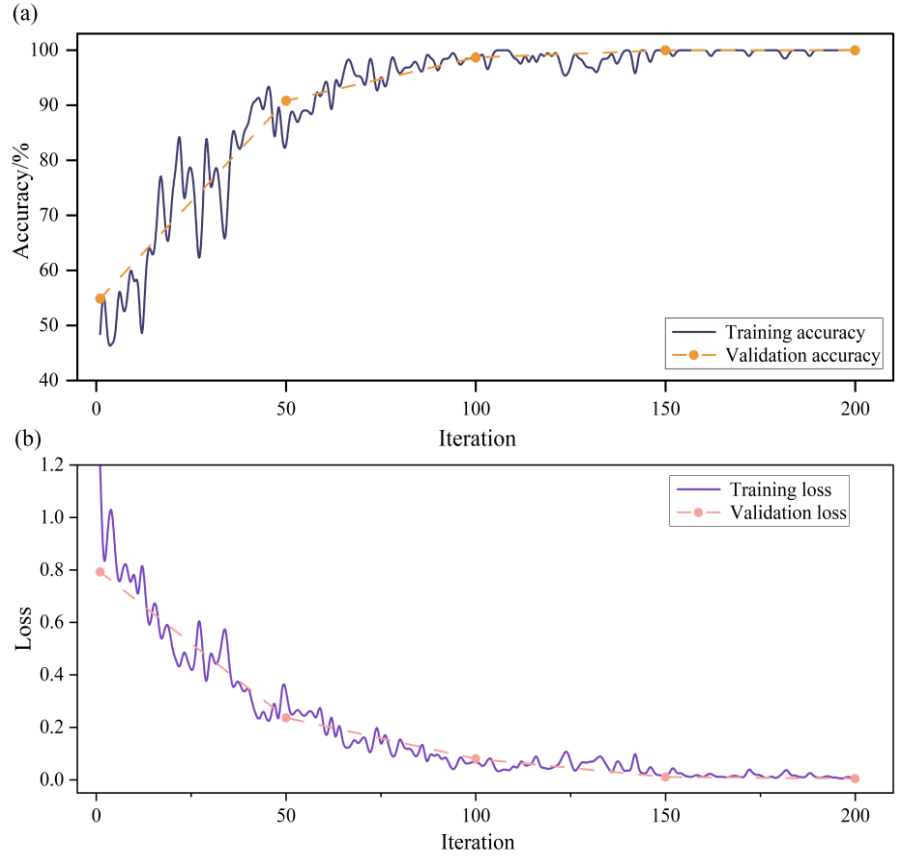

Fig. 16. Plots of CNN training and validation metrics: (a) Training accuracy and Validation accuracy; (b) Training loss and Validation loss.

As shown in Fig. 16a, it is found that the training set accuracy and validation set accuracy increases rapidly in about the first 100 iterations, and reaches nearly 100% at around the 150$th$ iteration and then remain stable, indicating that the model has achieved sufficient fitting to the data. Therefore, the training process is stopped at the 50th epoch (corresponding to 200 iterations) to prevent the model from overfitting. In Fig. 16b, both the training and validation loss values stabilize at approximately 0.01 and fully converge after around 150 iterations, which further confirms that the model has excellent stability.

To evaluate the performance of the proposed CNN model for image-based pipe defect classification, we calculate key performance metrics for each fold of the cross-validation, including the confusion matrix, as well as accuracy, precision, recall, and F1-score of each fold. Fig. 17 presents the 10 sets of test results (one set per fold) for each performance metric.

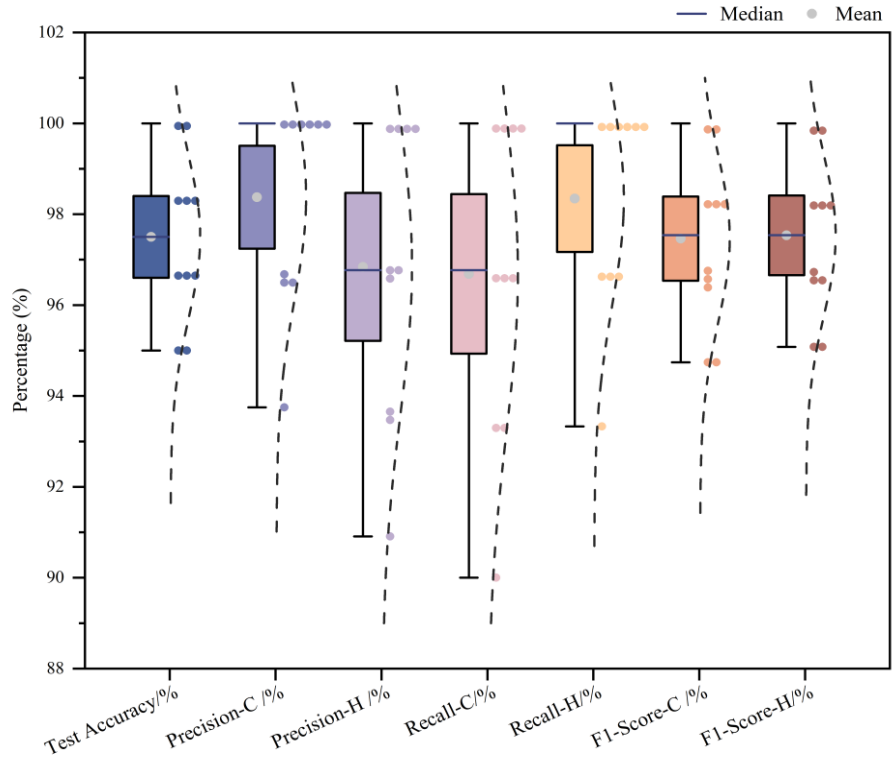

**Fig. 17**. Boxplots of classification performance metrics for the proposed CNN model's cross-validation tests.

Except for the medians of the crack precision (Precision-C) and hole recall (Recall-H) lying at the top of the boxplots, the medians of all other metrics are located around the center of their respective box; meanwhile, the mean values of all metrics also lie at the center of the box. This indicates that the data distributions of the performance metrics are symmetric and free of random contingency, thereby better reflecting the strong generalization ability of the proposed CNN model.

The F1-score integrates precision (ability to avoid false positives) and recall (ability to identify true positives), making it a robust metric for evaluating the comprehensive performance of a classification model. A high F1-score indicates that the model has achieved a fine balance between precision and recall. It is calculated that the average F1-scores of the hole and crack defect categories are 97.56% and 97.48%, respectively, with a difference of less than 0.1%, which demonstrates consistent classification performance of the model across different defect types. Additionally, the classification accuracy of both defect categories is evenly distributed between 95% and 100%, with an average value of 97.50%, and the variation among the 10 cross-validation results is not significant. The high F1-score and high classification accuracy not only collectively confirm the effectiveness of the proposed CNN model for defect classification, but also directly prove that high-precision classification results have been achieved.

**3.3.4 Performance comparison with other CNN architectures**

To evaluate the suitability of the proposed CNN model for image defect classification, its classification performance was compared with that of other popular deep learning architectures (such as AlexNet, DarkNet19, and VGG16). The comparison of classification performance between these four methods and the proposed method, in terms of validation accuracy, validation loss, test accuracy, Precision, Recall, and F1-score, is presented in Table 10. As shown in Table 10, the proposed network model achieves a test accuracy of 97.50%—the highest among all compared networks. Moreover, it outperforms the others in all other performance metrics. These results collectively confirm the advantages of the selected network.

Table 10. Comparison of classification performance metrics between the proposed network and other networks.

| Approach | Proposed CNN | AlexNet | DarkNet19 | VGG16 |
| --- | --- | --- | --- | --- |
| Validation Accuracy/% | 99.80 | 86.47 | 95.00 | 90.59 |
| Final Validation Loss | 0.0090 | 0.3302 | 0.2428 | 0.2627 |
| Test Accuracy/% | 97.50 | 86.50 | 96.00 | 89.17 |
| Precision-C/% | 98.37 | 91.08 | 97.67 | 94.05 |
| Precision-H/% | 96.84 | 87.18 | 94.60 | 82.74 |
| Recall-C/% | 96.69 | 84.33 | 94.33 | 83.00 |
| Recall-H/% | 98.34 | 88.67 | 97.67 | 93.57 |
| F1-score-C/% | 97.46 | 85.83 | 95.93 | 87.16 |
| F1-score-H/% | 97.53 | 86.72 | 96.07 | 89.50 |

## 4. Conclusions

This paper proposes an MVDR-CNN-based method for localization and classification of pipe defect. A uniform circular array is used to excite and receive UGW signals for localization via the MVDR algorithm. The constructed CNN model for image classification is trained using the Bayesian optimization and $k$-fold cross-validation method (with $k = 10$). The main conclusions are as follows: (1) Compared with the traditional pulse-echo method, the error of wave velocity in defective pipe measured via the CWT fitting method was reduced from 5.41% to 2.35%, which confirms the superiority of this method.

(2) For both holes and cracks, the average longitudinal and circumferential localization errors of the MVDR algorithm were both less than 1%. Meanwhile, no significant artifacts or side lobes were observed in the spatial spectrograms, which indicates that the algorithm has high resolution and can achieve accurate positioning. Compared with other estimation algorithms, MVDR algorithm offers certain advantages in terms of localization accuracy and efficiency.

(3) Regarding the two issues raised at the beginning of this paper—namely, high computational complexity and low SNR challenges—the time required for a single defect localization was only approximately 55 s, and the algorithm's positioning error was also less than 1% even under low SNRs. This demonstrates that the proposed algorithm can effectively address the aforementioned issues.

(4) The CNN model achieves a classification accuracy of 97.50% on the test dataset and performs well in terms of precision, recall, and F1-score, with average values all exceeding 97%. The proposed network model also outperforms other neural networks in classification performance metrics, which confirms the advantages of the proposed network.

Although the feasibility of this method is validated based on FE simulations, experiments should be conducted to examine the performance in industrial applications in future studies.

## Acknowledgements

Financial supports provided by the National Natural Science Foundation of China (Grant No. 12102145) and Natural Science Foundation of Jiangsu Province (Grant No. BK20210444) are acknowledged.

## References

[1] S. Zhang, J. Li, X. Wang, J. Zhang, Study on Horizontal Leakage Diffusion Characteristics of Small and Medium Aperture in High-Pressure Natural Gas Pipelines with Large Open Spaces, ACS Omega 10(10) (2025) 10515-10529.
[2] G. Xu, Z. Zhou, S. Xin, Z. Lin, L. Cai, Intelligent identification method for pipeline ultrasonic internal inspection, Nondestructive Testing and Evaluation 40(7) (2025) 2891-2912.

[3] X. Hong, Z. Luo, B. Zhang, G. Jin, Guided-wave quick sparse decomposition approach based on novel peak-frequency modulation dictionary for efficient composite plate damage inspection, Applied Acoustics 223(5) (2024) 110081.

[4] W. Shao, Y. Liao, y. Wang, J. Yin, G. Chen, X. Qing, A damage identification method for aviation structure integrating Lamb wave and deep learning with multi-dimensional feature fusion, Ultrasonics 151 (2025) 107623.

[5] J. Wei, L. Sun, C. Peng, L. Fan, F. Teng, W. Hao, L. Zhang, Q. Sui, M. Jiang, Probabilistic diagnostic imaging method based on the difference between the arrival moments of ultrasonic guided wave signals, Measurement 242 (2025) 116083.

[6] F. Teng, H. Zhang, S. Lv, L. Zhang, F. Zhang, M. Jiang, Improved-probabilistic imaging algorithm for separation problem of helical scattered path in pipeline ultrasonic guided wave inspection, Measurement 253 (2025) 117474.

[7] J. Wu, Y. Yang, Z. Lin, Y. Lin, Y. Wang, W. Zhang, H. Ma, Weak ultrasonic guided wave signal recognition based on one-dimensional convolutional neural network denoising autoencoder and its application to small defect detection in pipelines, Measurement 242 (2025) 116234.

[8] L. Li, X. Wang, X. Lan, T. Su, Y. Guo, A compressed 2D-DOA and polarization estimation algorithm for mmWave polarized massive MIMO systems, Digital Signal Processing 151 (2024) 104509.

[9] C. Mahapatra, A.R. Mohanty, Optimization of number of microphones and microphone spacing using time delay based multilateration approach for explosive sound source localization, Applied Acoustics 198 (2022) 108998.

[10] A. Fadakar, A. Jafari, P. Tavana, R. Jahani, S. Akhavan, Deep learning based 2D-DOA estimation using L-shaped arrays, Journal of the Franklin Institute 361(6) (2024) 106743.

[11] L. Chen, W. Mou, Z. Lv, Y. Zhang, L. Hu, G. Ou, Performance analysis of two-dimensional DOA estimation for uniform circular array, ICT Express 9(5) (2023) 854-859.

[12] R. Haeb-Umbach, T. Nakatani, M. Delcroix, C. Boeddeker, T. Ochiai, Microphone Array Signal Processing and Deep Learning for Speech Enhancement: Combining model-based and data-driven approaches to parameter estimation and filtering, IEEE Signal Processing Magazine 41(6) (2024) 12-23.

[13] L. Yu, M. Lyu, Y. Zhang, R. Wang, Y. Fang, W. Jiang, A hierarchical Dirichlet process for the background interference suppression to improve the microphone array imaging results, Mechanical Systems and Signal Processing 228 (2025) 112463.

[14] A. Mansourian, A. Fadakar, S. Akhavan, B. Maham, Robust 3-D Multi-Source Localization With a Movable Antenna Array via Sparse Signal Processing, IEEE Open Journal of the Communications Society 6 (2025) 3664-3682.

[15] Y. Han, J. Zhang, J. Luo, Relative DOA estimation method for UAV swarm based on phase difference information without fixed anchors, Scientific Reports 15 (2025) 14394.

[16] W. Wang, Y. Li, T. Shen, F. Liu, D. Zhao, An effective DOA estimation method for low SIR in small-size hydrophone array, Applied Acoustics 217(15) (2024) 109848.

[17] N.Z. Katz, M. Khatib, Y. Ben-Horin, J.D. Rosenblatt, T. Routtenberg, Geometry Design for DOA Estimation in Seismic 2D-Arrays: Simulation Study, IEEE Access 12 (2024) 35827-35843.

[18] C. Li, Z. Liu, L. Xie, B. Zhou, Q. Ma, Sparse reconstruction method for Direction-of-Arrival estimation using Vector sensor array, Signal Processing 238 (2026) 110176.

[19] J. Yin, K. Guo, X. Han, G. Yu, Fractional Fourier transform based underwater multi-targets

direction-of-arrival estimation using wideband linear chirps, Applied Acoustics 169 (2020) 107477.

[20] M.S. Bartlett, Smoothing Periodograms from Time-Series with Continuous Spectra, Nature 161(4096) (1948) 686-687.

[21] J. Capon, High-resolution frequency-wavenumber spectrum analysis, Proceedings of the IEEE 57(8) (1969) 1408-1418.

[22] E. Ortega, A. Vicente, A. Martínez, Ó. Rodríguez, M. Prieto, P. Parra, A. Da Silva, S. Sánchez, Enhancing efficiency in spaceborne phased array systems: MVDR algorithm and FPGA integration, Digital Signal Processing 155 (2024) 104732.

[23] C. Chen, G. Zhu, H. Sun, A class of high-resolution DOA estimation algorithms based on hyper-beamforming, Digital Signal Processing 158 (2025) 104921.

[24] Q. Zhou, Z. Wang, L. Huang, Q. Zhang, Y. Zhou, C. Yuan, A high-resolution DOA estimation via random forest virtual array extension, AEU - International Journal of Electronics and Communications 185 (2024) 155446.

[25] S. Jiang, S. Liu, M. Jin, High-dimensional MVDR beamforming based on a second unitary transformation, Signal Processing 205 (2023) 108869.

[26] Y. Zhu, Y. Zhang, F. Fan, W. Ma, L. Qin, Z. Kuang, M. Wu, J. Yang, An enhanced beamsteering algorithm based on MVDR for a multi-channel parametric array loudspeaker array, Journal of Sound and Vibration 595 (2025) 118768.

[27] L. Huang, Q. Zhou, S. Liao, B. Zhao, A high DOF and azimuth resolution beamforming via enhanced virtual aperture extension of joint linear prediction and inverse beamforming, Applied Acoustics 228 (2025) 110360.

[28] R. Schmidt, Multiple emitter location and signal parameter estimation, IEEE Transactions on Antennas and Propagation 34(3) (1986) 276-280.

[29] R. Roy, A. Paulraj, T. Kailath, ESPRIT--A subspace rotation approach to estimation of parameters of cisoids in noise, IEEE Transactions on Acoustics, Speech, and Signal Processing 34(5) (1986) 1340-1342.

[30] S. Zheng, Y. Luo, C. Xu, G. Xu, Combining circular laser sensing array with MUSIC algorithm for fast damage localization, Sensors and Actuators A: Physical 364 (2023) 114742.

[31] P. Raiguru, B.K. Swain, S.K. Rout, M. Sahani, R.K. Mishra, RDCSAE-RKRVFLN: A unified deep learning framework for robust and accurate DOA estimation, Applied Soft Computing 162 (2024) 111791.

[32] L. Wu, Y. Fu, X. Yang, L. Xu, S. Chen, Y. Zhang, J. Zhang, Research on the multi-signal DOA estimation based on ResNet with the attention module combined with beamforming (RAB-DOA), Applied Acoustics 231 (2025) 110541.

[33] M. Wang, X. Liu, X. Na, Y. Zhang, Y. Liu, T. Qiu, Neural network-based DOA estimation for distributed sources in massive MIMO systems, AEU - International Journal of Electronics and Communications 176 (2024) 155132.

[34] L. Xiao, G. Zheng, Y. Song, H. Zheng, Real-time and high-precision SVM-RF based prediction model for DOA estimation, Digital Signal Processing 161 (2025) 105122.

[35] L.T. Ramos, E. Casas, F. Rivas-Echeverría, Synthetic generated data for intelligent corrosion classification in oil and gas pipelines, Intelligent Systems with Applications 25 (2025) 200463.

[36] C. Jin, L. Zhou, Y. Pu, C. Zhang, H. Qi, Y. Zhao, Application of deep learning for high-throughput phenotyping of seed: a review, Artificial Intelligence Review 58 (2025) 76.

[37] B. Babu Vimala, S. Srinivasan, S.K. Mathivanan, Mahalakshmi, P. Jayagopal, G.T. Dalu, Detection and classification of brain tumor using hybrid deep learning models, Scientific Reports 13(1) (2023) 23029.

[38] S. Grigorescu, B. Trasnea, T. Cocias, G. Macesanu, A survey of deep learning techniques for autonomous driving, Journal of Field Robotics 37(3) (2020) 362-386.

[39] T. Nguyen-Da, P. Nguyen-Thanh, M.-Y. Cho, Real-time AIoT anomaly detection for industrial diesel generator based an efficient deep learning CNN-LSTM in industry 4.0, Internet of Things 27 (2024) 101280.

[40] P.N. Mueller, L. Woelfl, S. Can, Bridging the gap between AI and the industry — A study on bearing fault detection in PMSM-driven systems using CNN and inverter measurement, Engineering Applications of Artificial Intelligence 126 (2023) 106834.

[41] A. Jose, S. Shrivastava, An analytical examination of the performance assessment of CNN-LSTM architectures for state-of-health evaluation of lithium-ion batteries, Results in Engineering 27 (2025) 105825.

[42] N. Drir, F. Chekired, A. Mellit, N. Blasuttigh, Hybrid CNN-EML model for fault diagnosis in electroluminescence images of photovoltaic cells, Renewable Energy 250 (2025) 123343.

[43] M. Ertargin, A. Orhan, O. Yildirim, T. Gurgenc, Automated fault classification of asynchronous motor using mobile phone accelerometer and Parallel Residual CNN-GRU, Measurement 253 (2025) 117539.

[44] Z. He, Q. Liu, Deep Regression Neural Network for Industrial Surface Defect Detection, IEEE Access 8 (2020) 35583-35591.

[45] A. Arif, P.G. Rao, K. Prasad, A novel hybrid Bayesian-optimized CNN–SVM deep learning model for real-time surface roughness classification and prediction based on in-process machined surface image analysis, International Journal on Interactive Design and Manufacturing (IJIDeM) (2025).

[46] S. El-Hawwat, J.K. Shah, H. Wang, G. Venkiteela, Detection of internal cracks in polyethylene pipes using ultrasonic imaging and deep learning, Measurement 253 (2025) 117491.

[47] M. Gao, C.-T. Ng, J. Lin, A. Kotousov, Damage detection in circular tubes using nonlinear ultrasonic guided waves with metasurface, Thin-Walled Structures 214 (2025) 113384.

[48] K. Liu, Y. Fu, J. Ma, Multi-scale feature fusion based DOA and range estimation for near-field sources, Signal, Image and Video Processing 19(1) (2024) 4.

[49] L. Fan, X. Zhu, J. Zhu, Z. Xie, Z. Hu, M. Zhu, S. Xiong, Y. Ma, MVDR Beamforming Algorithm for Uniform Circular Array Based on Gain and Phase Errors, 2021 4th International Conference on Information Communication and Signal Processing (ICICSP), 2021, pp. 147-151.

[50] Z.İ. Aytaç, İ. İşeri, B. Dandil, A hybrid coot based CNN model for thyroid cancer detection, Computer Standards & Interfaces 94 (2025) 104018.

[51] Z. Zhao, E.B.A. Bakar, N.B.A. Razak, M.N. Akhtar, Assessment of Corrosion Image Rating Based on Transfer Learning, Arabian Journal for Science and Engineering (2024).

[52] F. Li, H. Zheng, X. Li, F. Yang, Day-ahead city natural gas load forecasting based on decomposition-fusion technique and diversified ensemble learning model, Applied Energy 303 (2021) 117623.

[53] Y. Zhang, Y. Yang, Cross-validation for selecting a model selection procedure, Journal of Econometrics 187(1) (2015) 95-112.

[54] S. Bates, T. Hastie, R. Tibshirani, Cross-Validation: What Does It Estimate and How Well Does It

Do It?, Journal of the American Statistical Association 119(546) (2024) 1434-1445.

[55] H. Chen, Z. Zhang, W. Yin, C. Zhao, F. Wang, Y. Li, A study on depth classification of defects by machine learning based on hyper-parameter search, Measurement 189 (2022) 110660.

[56] S.M. Malakouti, Improving the prediction of wind speed and power production of SCADA system with ensemble method and 10-fold cross-validation, Case Studies in Chemical and Environmental Engineering 8 (2023) 100351.

[57] Z. Xu, X. Zhang, A stochastic configuration network based on an improved sampling strategy and Bayesian optimization, Knowledge-Based Systems (2025) 113879.

[58] S. Azadi, Y. Okabe, V. Carvelli, Bayesian-optimized 1D-CNN for delamination classification in CFRP laminates using raw ultrasonic guided waves, Composites Science and Technology 264 (2025) 111101.